# Bulk Electronic Structure of Lanthanum Hexaboride (LaB$_6$) by Hard X-ray Angle-Resolved Photoelectron Spectroscopy

Arunothai Rattanachata,[1,2,3] Laurent C. Nicolaï,[4] Henrique P. Martins,[1,5] Giuseppina Conti,[1,5] Matthieu J. Verstraete,[6] Mathias Gehlmann,[1,2] Shigenori Ueda,[7,8] Keisuke Kobayashi,[9] Inna Vishik,[1] Claus M. Schneider,[1,2,10] Charles. S. Fadley,[1,2,†] Alexander X. Gray,[11] Jan Minár,[4] and Slavomír Nemšák[5]

[1]*Department of Physics, University of California, Davis, California 95616, USA*
[2]*Materials Sciences Division, Lawrence Berkeley National Laboratory, Berkeley, California 94720, USA*
[3]*Synchrotron Light Research Institute, Nakhon Ratchasima 30000, Thailand*
[4]*New Technologies-Research Center, University of West Bohemia, 306 14 Pilsen, Czech Republic*
[5]*Advanced Light Source, Lawrence Berkeley National Laboratory, Berkeley, California 94720, USA*
[6]*Nanomat Q-MAT CESAM and European Theoretical Spectroscopy Facility, University of Liege, 4000 Liege, Belgium*
[7] *Research Center for Functional Materials, NIMS, 1-1 Namiki, Tsukuba, Ibaraki 305-0044, Japan*
[8]*Synchrotron X-Ray Station at Spring-8, National Institute for Materials Science (NIMS), 1-1-1 Kouto, Sayo, Hyogo 679-5148, Japan*
[9]*Quantum Beam Science Directorate, Japan Atomic Energy Agency, 1-1-1 Kouto, Sayo, Hyogo 679-5148, Japan*
[10]*Peter-Grünberg-Institut PGI-6, Forschungszentrum Jülich, 52425 Jülich, Germany*
[11]*Department of Physics, Temple University, Philadelphia, Pennsylvania 19122, USA*

*E-mail addresses: a.rattanachata@gmail.com; grconti@lbl.gov; snemsak@lbl.gov

**Abstract**

In the last decade rare-earth hexaborides have been investigated for their fundamental importance in condensed matter physics, and for their applications in advanced technological fields. Among these compounds, LaB$_6$ has a special place, being a traditional d-band metal without additional f- bands. In this paper we investigate the bulk electronic structure of LaB$_6$ using hard x-ray photoemission spectroscopy, measuring both core-level and angle-resolved valence-band spectra. By comparing La 3d core level spectra to cluster model calculations, we identify well-screened peak residing at a lower binding energy compared to the main poorly-screened peak; the relative intensity between these peaks depends on how strong the hybridization is between La and B atoms. We show that the recoil effect, negligible in the soft x-ray regime, becomes prominent at higher kinetic energies for lighter elements, such as boron, but is still negligible for heavy elements, such as lanthanum. In addition, we report the bulk-like band structure of $LaB_6$ determined by hard x-ray angle-resolved photoemission spectroscopy (HARPES). We interpret HARPES experimental results by the free-electron final-state calculations and by the more precise one-step photoemission theory including matrix element and phonon excitation effects. In addition, we consider the nature and the magnitude of phonon excitations in HARPES experimental data measured at different temperatures and excitation energies. We demonstrate that one step theory of photoemission and HARPES experiments provide, at present, the only approach capable of probing true "bulk-like" electronic band structure of rare-earth hexaborides and strongly correlated materials.

---

[†] Deceased August 1$^{st}$ 2019.

## I. Introduction

Hexaboride compounds have been extensively studied in the last thirty years.[1,2] Among these compounds, lanthanum hexaboride ($LaB_6$), a thermionic material, has been studied for its high electron emissivity and very low work function of ~2.5 eV for the (100) surface orientation.[3] It is one of the most widely-used materials for hot cathodes in electron microscopy and lithography. Lately, it has been studied as a promising candidate for solar energy, photonic and electronic applications.[4,5,6] Among the class of rare-earth hexaboride compounds, $LaB_6$ has a special place, being a traditional d-band metal without additional f- bands. Other rare-earth hexaborides ($RB_6$) are investigated for their intrinsic fundamental importance in condensed matter, and for their applications in advanced technological fields.[7] For example, $CeB_6$ is a dense Kondo material that shows electric quadrupole ordering,[8,9] $EuB_6$ is a ferromagnetic semi-metal,[10,11] $YbB_6$ is a fairly correlated $Z_2$ topological insulator.[12] $SmB_6$ contains additional f bands which hybridize with the d-electrons to produce a narrow-gap Kondo insulator.[13,14,15,16] At present, several research groups are debating if $SmB_6$ could be also a topological Kondo insulator.[17] However, due to the narrow band gap, the in-gap edge states near the Fermi level are very difficult to detect.[18] The theoretical prediction that $SmB_6$ could be a topological Kondo insulator led to renewed and intense scientific interest in the full class of rare-earth hexaborides.[19]

The surface electronic structure of these materials has been widely studied both theoretically[20,21,22,23] and experimentally.[6,15] Over the past 50 years, angle-resolved photoelectron spectroscopy (ARPES), has developed into a powerful technique for determining the electronic structures of crystalline materials.[24,25] Although ARPES has played an important role in condensed-matter physics research, its capability has been limited to the characterization of surfaces, due to the short electron inelastic mean free paths (IMFP) in solids. In a typical ARPES experiment with photon energies ranging from ~10 to 100 eV, the probing depth is about ~10 Å and surface effects are dominant, in particular for rare-earth compounds with strong electron correlations.

In order to overcome this limitation, in the last decade considerable effort has been invested in the use of x-ray energies ranging from the sub-keV (soft) to multi-keV (tender/hard x-ray regime) in order to achieve higher sampling depth and to better study the properties of the bulk.[26,27,28] Previously, hard X-ray ARPES (HARPES) was established as a technique studying dilute magnetic semiconductors.[29,30,31] By using these higher excitation energies, it is possible to overcome the surface sensitivity limitations of ARPES and measure the experimental $LaB_6$ bulk electronic structure.

We studied $LaB_6$ by tender/hard x-ray photoelectron spectroscopy (XPS), measuring core levels and valence band spectra, and by hard x-ray angle-resolved photoemission spectroscopy (HARPES). Beyond being an archetypal hexaboride, $LaB_6$ is a very interesting case-study for hard XPS because it is a stoichiometric compound with a large mass difference between heavy La (atomic weight 138.9 u) and light B (atomic weight 10.8 u). Thus, within one crystal we have two elements that will react very differently to the hard x-ray excitation: we expect to see different values of recoil energy from lanthanum and boron core levels, and different behaviors of the dispersive valence bands in HARPES due to thermal vibrations and the phonon creation and annihilation processes.

In this paper, we also demonstrate a powerful synergy between one step theory of photoemission and HARPES experiments in interpreting the measured data. This combination of techniques provides, at present, the only approach capable of probing, both experimentally and theoretically, true "bulk-like" electronic band structure of materials

## II. Methods

The sample we investigated is a commercially available single-crystal $LaB_6$ (001) from Kimball Physics Inc. The XPS and HARPES experiments were performed at the synchrotron radiation facility SPring-8 in Hyogo Japan, at the undulator beamline BL15XU, and at the Advance Light Source (ALS) at Berkeley USA, at the bending magnet beamline BL 9.3.1. For the data collected at BL15XU, we used x-ray photon energies of hν = 3237.5 eV and hν = 5953.4 eV with total instrumental resolutions of 180 meV and 240 meV, respectively. For the data collected at BL 9.3.1, we used a photon energy of hν = 2830 eV with total instrumental resolution of about 600 meV. Figure 1 shows the experimental geometry for both experiments. The exciting radiation was incident on the sample at the grazing angle of 2.0° measured from the sample surface plane. The photo-emitted electrons were collected and analyzed for their kinetic energy by VG Scienta R4000 (BL15XU) and Scienta SES 2002 (BL 9.3.1) spectrometers. The sample was cooled by liquid helium down to ≈ 30 K at BL15XU and by liquid nitrogen down to ≈ 90 K at BL 9.3.1. The inset in Figure 1 shows a schematic of a unit cell of the $LaB_6$ crystal structure. The binding energies were calibrated by measuring the Fermi edge and the Au 4f core level of an Au standard sample.

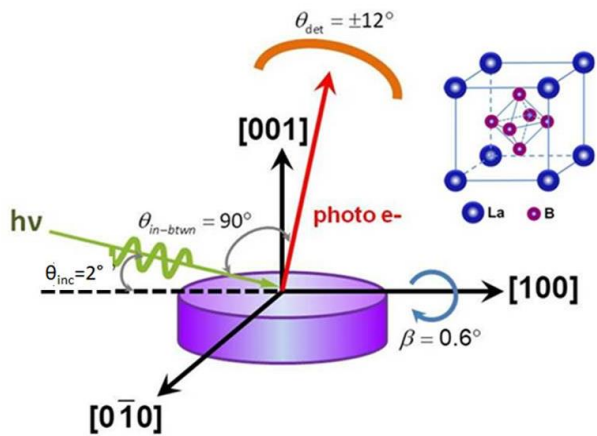

**Figure 1.** Experimental geometry used at BL15XU, Spring-8. The inset shows the crystal structure of $LaB_6$. The crystal structure of $LaB_6$ is simple cubic (space group Pm-3m), with $B_6$-octahedra in body-centered positions and La atoms at the corners of the unit cell.

The radiation was p-polarized, with the degree of linear polarization being ≈0.9 at 3237.5 eV and ≈1.0 at 5953.4 eV, with the polarization vector pointing in the direction of the analyzer. The electron takeoff angle was 88° as measured from the sample surface plane, which maximizes bulk sensitivity. This experimental geometry yields an x-ray attenuation length of 0.121 μm at 3237.5 eV and of 0.153 μm at 5953.4 eV,[32] which ensures that the x-rays penetrate deep into the bulk. Information depth of the photoemission experiment, which is proportional to inelastic mean free path (IMFP) of valence electrons, was estimated using the TPP-2M formula.[33] The calculated IMFP for kinetic energies of 3237.5 eV and 5953.4 eV are 51.63 Å and 85.59 Å, respectively. Thus, the average probing depth of the valence band measurements at these two energies corresponds to about 12.4 unit cells (u.c.) and 20.6 u.c. (lattice constant a = 4.15597 Å)[34], respectively, providing truly bulk-sensitive measurements of the electronic properties. Previously, only studies performed in the soft x-ray regime were reported in literature at hν = 70 eV and hν = 126 eV, corresponding to ≈ 1 u.c. and 1.4 u.c. of average proving depth.[15,35]

In order to support of the experimental data, ab-initio calculations using the SPR-KKR[36] package were performed. This package is based on the Dirac equation thus fully accounting relativistic effects. More specifically, the Local Density Approximation (LDA) was used in order to derive the crystal potential. Nevertheless, a step further is required for more pertinent comparison with experimental measurements. Such theoretical calculations were obtained using the one-step model of photoemission which describes accurately the excitation process, the transport of the photoelectron to the crystal surface as well as the escape into the vacuum as a single quantum-mechanically coherent process including all multiple-scattering events. This one-step model of photoemission incorporates surface and matrix element effects and therefore allows a thorough, quantitative comparison to the experimental HARPES data. The displacements due to the final temperature were included by means of a Debye-Waller (DW) factor introduced by a Coherent Potential Approximation (CPA) analogy.[37,38,39] In addition, in order to go beyond the Debye-Waller approximation, the chemical element specific phonon contributions were further taken into account by including Density Functional Perturbation Theory (DFPT) derived mean displacements (see Figure 8a and Supplemental Material (SM-4). The phonons of $LaB_6$ were calculated using DFPT[40,41,42] as implemented in the ABINIT software package.[43] The DFT calculations used norm conserving pseudopotentials of the ONCVPSP form from the pseudo-dojo web site[44] the GGA-PBE [45] approximation for the exchange correlation energy, a plane wave basis set cutoff of 45 Ha, a 12x12x12 grid of wave vectors for the Brillouin zone sampling of electrons and 4x4x4 for the phonons. The total energy was tightly converged to $10^{-13}$ Ha for the ground state and $10^{-9}$ Ha for the phonon perturbations. The stress was relaxed to within less than $10^{-2}$ GPa and forces below $2 \times 10^{-5}$ Ha/Bohr (Ha/Bohr is 51.4221 eV/Å).

The harmonic approximation was used with a fixed volume (a = 4.15597 Å): thermal expansion has a negligible effect at least up to room temperature yielding phonon dispersion relations in good agreement with the literature, as well as thermodynamic quantities, and the atomic mean square displacements (MSQD) as a function of temperature. The MSQD were incorporated in the full KKR as described above.

## III. Experimental results

### a) Chemical Analysis

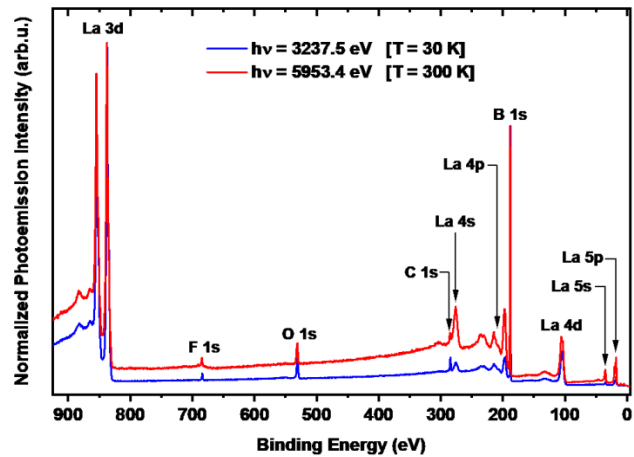

**Figure 2.** Survey spectra of LaB$_6$, at 3237.5 eV (blue) and at 5953.4 eV (red) photon energies. The presence of O, C and F atoms suggests surface contamination.

Figure 2 shows the survey spectra measured at photon energies of 3237.5 eV and 5953.4 eV. All major features are assigned to core-levels (CL) of La and B as expected. In addition, we observed C 1s and O 1s signals, which indicate some sample contamination and F 1s originating from a residual chamber contamination.

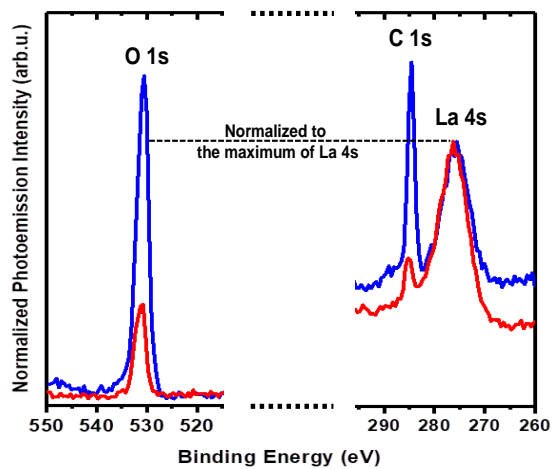

**Figure 3.** Core level spectra O 1s and C 1s normalized against La 4s measured at photon energies, 3237.5 eV (blue plot) and 5953.4 eV (red plot), respectively.

In order to verify whether C 1s and O 1s are contaminants on the sample surface or in the "bulk", we measured these CL in detail (Figure 3) and we normalized their intensities against the La 4s peak. Figure 3 shows that these normalized intensities are higher in the spectrum collected at the lower photon energy (3237.5 eV) compared to that collected at higher photon energy (5953.4 eV); this is a clear evidence that carbon and oxygen contamination are closer to the sample surface. The thickness of this surface contamination layer was modeled using the NIST database program Simulation of Electron Spectra for Surface Analysis (SESSA).[46] Comparing the experimental intensities of C 1s and O 1s to those of the simulation, the thickness of this surface contamination was determined to be about 10 Å where the estimated density is $5\times10^{22}$ atoms/cm$^3$. In addition, comparing the areas of the B 1s and La 5p peaks, corrected by their relative photoionization cross-section, to the areas obtained by SESSA, we conclude that our LaB$_6$ sample is stoichiometric to within 3%, which lies within the expected accuracy of the XPS analysis, using tabulated photoionization cross-sections and analyzer transmission function. In addition we observed that the binding energy of the B 1s peak (figures 2 and 5a) is close to that of a pure boron compound, and that no other peaks due to possible Boron oxidation states are detected.[47] However, we noticed that the binding energy of B 1s shows a shift with photon excitation energies. This shift in binding energy is due to the recoil effect and will be discussed in detail later.

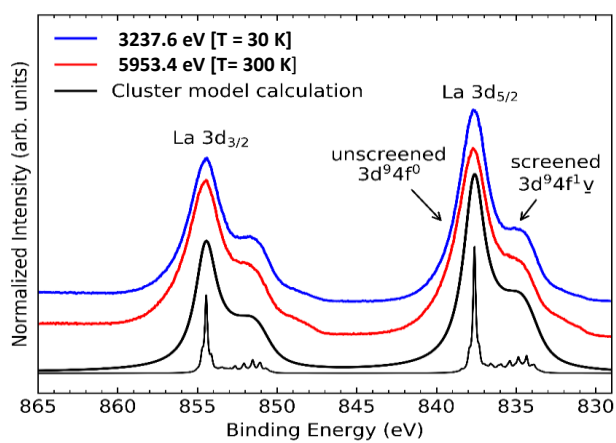

**Figure 4.** La 3d core level photoemission spectra of LaB$_6$ measured at photon energies of 3237.5 eV (blue line) and 5953.4 eV (red line) compared to the calculated spectral weight of a cluster model calculation (thick black line is convoluted with a Voigt function, while the thin black line is the calculated unbroadened spectral weight). The calculated spectrum shows the final state effect of the photoemission core-hole that leads to the poorly-screened and well-screened peaks in the experimental spectra.

Figure 4 shows the more complex La 3d region measured at photon energies of 3237.5 eV and 5953.4 eV compared to cluster model calculations as described in the following section. The La 3d spin-orbit splitting doublet is at 837.9 eV for La 3d$_{5/2}$ and at 854.8 eV for La 3d$_{3/2}$. The separation between these peaks is about 17 eV in agreement with literature for non-oxidized lanthanum. Each of these two main peaks is accompanied by a satellite with lower intensity and a separation of about 3 eV from the main peaks. Based on the B 1s and La 3d binding energies and intensities we can conclude that our LaB$_6$ sample does not show detectable presence of B-O$_x$ or La-O$_x$ in the bulk.[48]

**b) Orbital Hybridization**

Since the early seventies, the doubling of each of the La 3d$_{3/2}$ and La 3d$_{5/2}$ peaks was extensively investigated in numerous studies of insulating[49,50] as well as metallic[51,52] compounds, and attributed to final-state screening effects.[53,54] Similar to La 3d, La 4d core levels also show such satellite structures; however, because the main La 4d spin-orbit splitting is only 2.7 eV, the satellite peaks are not as clearly resolved, and produce a more complex shape of the spectrum.[55] As a consequence, in this paper, we focus our discussion only on the La 3d region.

When interpreting core level spectra, the main peak is usually attributed to the well-screened final states, while the poorly-screened satellite appears at higher binding energies. However, this general assumption is not always correct for transition metal based compounds: the satellite can also appear at lower binding energies than the main peak. In order to interpret these final-state satellites, Kotani and Toyozawa[56] proposed a model in 1974 which is still accepted today. In the ground state, the La 3d orbitals are full, and the La 4f level is empty, well above the Fermi level (E$_F$). In the final state of the photoemission process, the La 4f level on the core-hole site is assumed to be pulled down below E$_F$ because of the attractive core-hole potential. According to this mechanism, the final state splits into two configurations: one where, through hybridization, an electron near the E$_F$ populates the 4f level (3d$^9$4f$^1$v̲ configuration, where v̲ denotes a hole in the valence band) and screens the core-hole potential, the other where the 4f level remains empty (for the duration of the photoemission event) even after its energy has been lowered below the original E$_F$ (3d$^9$4f$^0$ configuration).

In order to correctly assign the La 3d bands, we performed configuration interaction cluster model calculations where charge-transfer and final-state effects due to core-hole screening are included.

Figure 4 shows the La 3d photoemission spectrum compared to cluster model calculations. The model consists of the La 3d core level, the La 4f level, and a filled valence band, and takes into account local 3d-4f electron-electron interactions and hybridization effects. The Hartree-Fock values of F$^k$ and G$^k$ Slater integrals that describe the atomic 4f-4f and 3d-4f electron interactions were reduced to 80% to account for bulk intra-atomic screening effects.[57] The exchange integrals and spin-orbit splitting values for the initial and final states were calculated using the Cowan code.[58] The parameters of the model are the charge transfer energy $\Delta = E(4f^1\underline{v})$ -$E(4f^0)$, the effective hybridization $V_{eff}$ between the La 4f and valence band states and the 3d-4f core-hole attractive potential $U_{fc}$. For the La 3d photoemission spectrum of LaB$_6$ these parameters were fixed to $\Delta = 11$ eV, $V_{eff} = 0.32$ eV, and $U_{fc} = 12.6$ eV. The valence band was described by a discrete set of N levels with a band-width of w = 2.5 eV; the number of N levels was increased until convergence was achieved at N = 4. The calculations were performed using the Quanty package[59] and the resulting spectral weight is obtained through a Green's function formalism. The calculated unbroadened and broadened spectral weights are shown at the bottom of Figure 4. The broadened curve was calculated by convoluting the spectral weight with a Voigt profile to account for experimental resolution and core-hole lifetime effects. Since $\Delta$ is relatively large and the hybridization between the valence band and the La 4f states is small, the calculated ground state of the system shows a mostly empty La 4f level, as expected, and is composed of 98.8% $4f^0$ and 1.2% $4f^1\underline{v}$ configurations. In the final states, however, the La 4f states are pulled down by the core-hole potential which leads to an effective charge-transfer energy $\Delta_f = \Delta - U_{fc} = -1.6\,eV$ The case where the 4f level remains empty, after the core-hole is created, leads to the main poorly-screened peak ($3d^{10}4f^0 \rightarrow 3d^94f^0$ transitions), whereas, if an electron near the $E_F$ screens the core-hole, it gives rise to the well-screened satellite peak dominated by $3d^{10}4f^1\underline{v} \rightarrow 3d^94f^1\underline{v}$ configuration. The negative value of $\Delta_f$ increases the contribution of well-screened $3d^94f^1\underline{v}$ states in the possible final states, even if the hybridization is small. As a result, this well-screened contribution appears at a lower binding energy compared to the main poorly-screened peak, as can be seen in Figure 4. The relative intensity of the La 3d two peaks depends on the hybridization between the La 4f and the valence electrons. In case of pure La metal, the intensity of the La 3d well-screened peak is very small because the hybridization is almost negligible.[52] In LaB$_6$, instead, there is a small but non-negligible hybridization between La 4f orbitals and B s-p bands. As a consequence, clear well-screened satellites ($\approx 835$ eV and $\approx 852$ eV) are visible in the LaB$_6$ XPS spectrum (Figure 4).

**c) Recoil Effect**

We have observed that the binding energy of B 1s core-level increases with higher excitation energy. None of the La core levels exhibit any detectable shift, therefore, the B 1s peak shift is not due to effects such as charging, surface oxidation, or instrumental instability, and we attribute it to the recoil effect.[60,61]

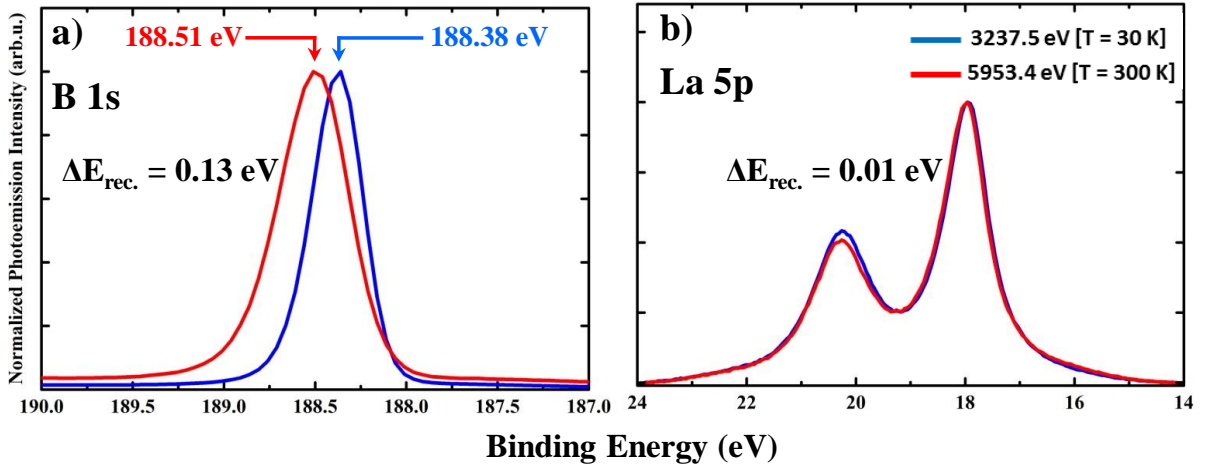

**Figure 5**. XPS spectra a) B 1s and b) La 5p measured at photon energies, 3237.5 eV (blue plot) and 5953.4 eV (red plot), respectively.

The recoil effect was theoretically predicted in 1976[62,63] but was not observed experimentally until 2007, when high brilliance third-generation synchrotron radiation sources became available for hard XPS.[27] It was first observed by Takata el al.[64] for the C 1s core level of graphite and later in the valence band region for Al[65] and then for Cr and $Cr_{1-x}Al_x$ alloys by Boekelheide et al..[66] The recoil energy is negligible in traditional soft x-ray photoemission experiments, but if the kinetic energy of the emitted electron is high enough, which can occur in hard x-ray photoemission, the recoil effect becomes visible in the data.

As the photoelectron is ejected from an atom, part of the photon energy transforms into motion of the atom due to the conservation of momentum. The loss of the electron kinetic energy $E_{Recoil,i}$ after ejection from an atom *i* can be quantified as

$$E_{\text{Recoil},i} \approx \frac{E_{kin.} m_e}{M} \qquad (1)$$

where $E_{kin}$ is the initial kinetic energy of the photoelectron, $m_e$ is the mass of the electron and $M$ is the mass of the atom. The quantity $E_{Recoil,i}$ is called the recoil energy. Note that Eq.(1) neglects the effect of the emitting atom being bound to a lattice.

In this work we compare the recoil effects in the $LaB_6$ core electrons at two different excitation energies hν = 3237.5 eV and hν = 5953.4 eV. $LaB_6$ is a very interesting case-study for the recoil effect in hard XPS, since it consists of a heavy atom (La, atomic weight 138.9 u) and a light atom (B, atomic weight 10.8 u) in the same compound. Their recoil energies will be substantially different, much greater for B than for La. Figure 5 shows B 1s, and La 5p core levels collected at those two photon energies. The B 1s and La 5p core levels did not show intrinsic peak broadening, the bandwidth variation in B 1s at the two different photon energies is consistent with the corresponding total instrumental energy resolution. Shirley background was subtracted[67] from these spectra and their heights were normalized to their peak maximum.

In Figure 5a, a difference of 0.13 eV in the binding energies of the B 1s peak is observed at those two photon energies. The calculated difference using

$$\Delta E_{\text{Recoil},B} = E_{\text{Recoil},B}(5953.4\,eV) - E_{\text{Recoil},B}(3237.5\,eV) \qquad (2)$$

yields 0.14 eV for B 1s. Figure 5b instead shows that La 5p$_{3/2}$ does not exhibit any detectable shift, and the calculated value of $\Delta E_{\text{Recoil},La}$ is 0.01 eV, too small to be observed in our experiment.

The agreement between the experimental recoil shifts with the calculated value from Eq. (1), which is valid for emission from an isolated atom, suggests, in the first approximation, that the boron atoms have little interaction with the rest of the atomic lattice during the photoemission process.

The presence of core-level recoil energy shifts of various magnitudes is an important aspect in hard x-ray photoemission spectroscopy, and has to be carefully taken into account in the interpretation of the data. Furthermore, the recoil energy is intimately linked to the photon-phonon interaction and to the fraction of transitions in which the entire lattice could recoil, as described by the DW factor. Due to this interaction, the La and B dispersive valence bands will show different behavior in the HARPES data. For instance, the creation and annihilation of phonons during photoemission can cause a smearing along the momentum axis in the HARPES data, as discussed in detail below.

**d) Valence-band structure**

With the development of third-generation, and the dawn of fourth-generation synchrotron sources, [28] HARPES is now being used routinely to probe the "true" bulk properties of solids, providing greater information depth due to the large inelastic mean-free-path of the escaping photoelectrons.[68,69,31] However, ARPES measurements in the hard x-ray regime present specific challenges due to quite low photoionization cross sections, strong phonon scattering, photoelectron diffraction effects, and large photon momentum transfer. We report in this section HARPES applied to LaB$_6$ where, as calculated from IMFP, about 12 u.c. and about 21 u.c, are probed at excitation energy $h\nu = 3237.5$ eV and at $h\nu = 5953.4$ eV respectively.

In addition to the ARPES data collected at $hv$ = 3237.5 eV eV and $T$= 30 K, shown in Figure 6, we have collected ARPES data at $hv$ = 2830 eV and $T \approx$ 90 K as well as at $hv$ = 5953.4 eV and $T$ = 300 K. The experimental ARPES images of these other two experiments are reported in Supplemental Material SM-1.

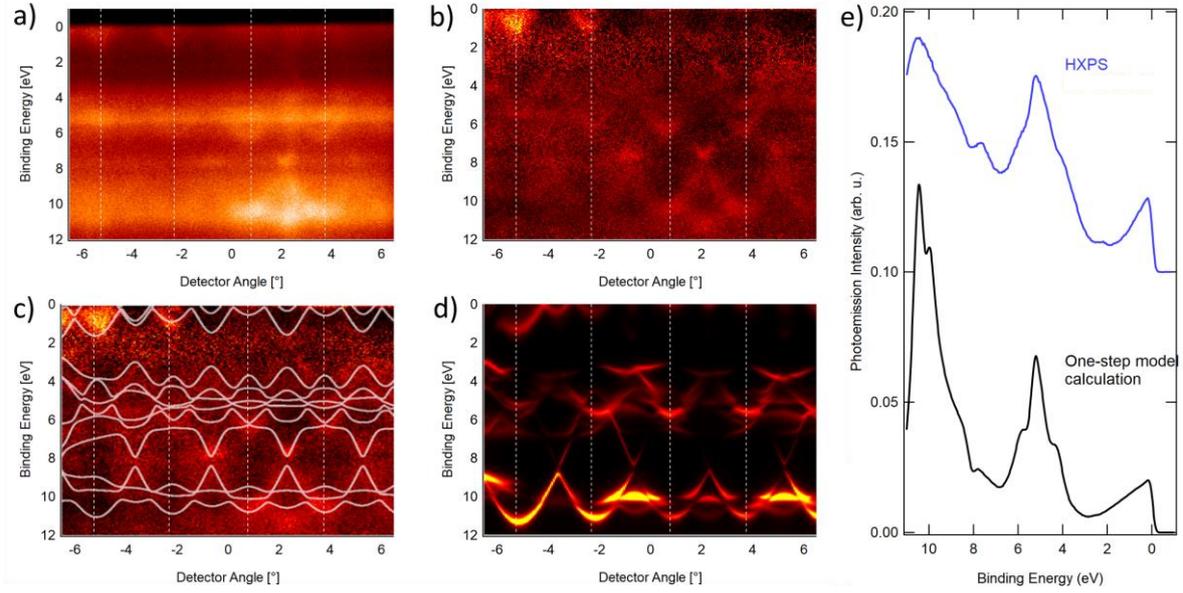

**Figure 6.** a) ARPES raw image collected at $hv$ = 3237.5 eV and temperature 30 K; Fermi Energy is at BE = 0 eV; b) ARPES image normalized by the density-of-states (DOS) and by the x-ray photoelectron diffraction (XPD) modulations; c) Free Electron Final State (FEFS) band structure calculation (gray solid lines) overlaid on the ARPES data from b); d) angle-resolved photoemission data calculated by the one-step theory at $hv$ = 3238 eV and temperature 30 K; e) experimental (HXPS at $hv$ = 3237.5 eV, $T$ = 30 K ) and calculated ($hv$ = 3238 eV, $T$ = 30 K) angle-integrated valence band intensities

Figure 6a shows the detector image of LaB$_6$ ARPES, measured at the photon energy of 3237.5 eV and at a temperature of 30 K. No cleaning of the atmosphere-exposed LaB$_6$ surface was done. Cooling the sample allows to partly suppress the smearing of the band structures, which is due to phonon-assisted indirect transitions. The phonon smearing can be estimated using a photoemission DW factor $W(T)$, which predicts the fraction of the direct transitions for a given photon energy and temperature calculated from

$$W(T) = \exp[-\frac{g_{hkl}^2}{3} <U^2(T)>] \qquad (3)$$

where $g_{hkl}$ is the magnitude of the bulk reciprocal lattice vector involved in the direct transitions for a given photon energy and $U^2(T)$ is the element-specific three dimensional mean-squared vibrational displacement for a given temperature.[70,71,29]

To estimate the Debye temperature of LaB$_6$, M.M. Korsukova at al.[72] considered that the two La and B sub-lattices have different Debye temperatures: the more loosely bound lanthanum ions are assigned a Debye temperature of 417 K and the rigid boron network to 732 K. This assumes relatively weak coupling between the two sub-lattices. Based on the *W(T)* as reported in [72] and on Eq.(3) we estimated that the DW factor for LaB$_6$ is about 0.6 at 3237.5 eV and at 30 K, as a consequence, we expect to observe a clear bulk dispersive band structure.

The detector image in Figure 6a is dominated by the density-of-states (DOS) and by the x-ray photoelectron diffraction (XPD) modulations. The effects of DOS (Eq. 4a) and XPD (Eq. 4b) can be removed by normalizing this image by the integrated photoelectron intensities:

$$DOS \approx \int I(E,\vartheta)d\vartheta \qquad (4a)$$

and

$$XPD \approx \int I(E,\vartheta)dE \qquad (4b)$$

This normalization process also removes the matrix element effects. Figure 6b shows the normalized enhanced angle-resolved bulk band electronic structure corrected also by the relative angular shift of the band structure features with respect to the x-ray photoelectron diffraction pattern. We calculated that this relative angular shift in our experiment is 3.219° (see Supplemental Material SM-2).

## IV.  Theoretical calculations

### a) Valence band photoemission calculations

We compared the experimental ARPES data reported in Figure 6b to bulk electronic band structure calculated by DFT using free-electron final-state (FEFS) approximation and by the more accurate one step model of photoemission, including lattice vibrations.[36] FEFS calculations using DFT generated band structures are reported in Figure 6c. In Supplemental Material note SM-3 we describe the FEFS model. A particular challenge in HARPES experiments, compared to conventional low-energy ARPES, is the extreme sensitivity to the sample alignment. We found that the best agreement between experiment and calculation is given with an additional tilt of 0.6° compared to the initially assumed sample geometry (SM-3).[73] Such a small misalignment is caused by imperfect mounting of the sample to the sample holder and it is to be expected. Once the optimized geometries were obtained using the FEFS model, we proceeded to simulate the HARPES spectra with the one-step photoemission model.[74,75]

Unlike the FEFS calculations, the one-step calculations include the concepts of light polarization, surfaces and final-states effects, thus allowing quantitative comparison to the experimental HARPES data. Figure 6d shows the calculated band dispersions results from the one step theory, with white dashed lines being guides to the eye, marking neighboring Brillouin zones. The agreement between the features present in the experimental ARPES data and the theoretical calculations is remarkable. In addition, as a further more quantitative comparison between the experimental and calculated valence band photoemission spectra, we calculated angle-integrated counterparts of angle-resolved data shown in Figures 6a and 6d. These results are shown in Figure 6e, with all the features present in the experimental data also present in the calculated spectra, with slightly better resolution in the calculated spectra. Calculated spectra were convoluted with 250 meV wide Gaussian function to account for the experimental broadening.

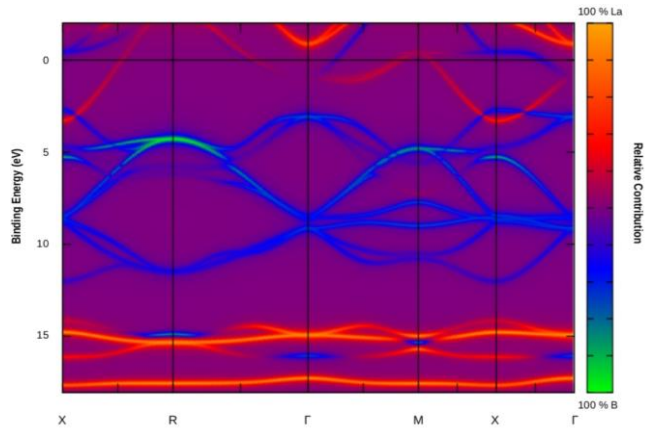

**Figure 7.** Calculated Bloch spectral functions projected onto La and B sites, showing mixed character of bands throughout the valence band region.

As a next step, we calculated the theoretical projected Bloch spectral functions (BSF) onto La and B sites along certain high symmetry orientations, as shown in Figure 7. The BSF plot clearly shows a mixed character of bands throughout the entire valence band region. B contribution is mostly between 5 eV and 12.5 eV, although La is not negligible. The La contribution, on the other hand, is mostly concentrated between 0 and 4 eV and below 12.5 eV. The character of the individual bands will be discussed with regards to phonon excitations, as described in the next section.

b) **Vibrational calculations**

Figure SM-4a shows the phonon band structure of $LaB_6$ as calculated by DFPT (see Methods section), with element projected Density of States (DOS) in side panel. The optical

manifold is completely dominated by B and separated by an energy gap from the acoustic manifold which moves only the La atom. The Debye model is inappropriate in both cases: for B the modes are rarely linear in dispersion except around 50 meV and for La much of the dispersion is flat beyond the first fifth of the Brillouin Zone. Figure SM-4b shows the specific heat of $LaB_6$ within the harmonic approximation and DFPT, as a function of temperature. The inset shows the small bump created by the lanthanum manifold around 50 K. Figure 8a shows the mean square displacement of the La and B atoms, as a function of T. The light B atoms have significant zero-point motion contributions at T = 0 K, but increase more slowly as they have no low frequency contributions. The heavy La atoms start with a smaller zero point effect, but increase more quickly with T due to the increasing Bose Einstein occupation of the acoustic modes. The complex phonon band structure of $LaB_6$ explains the contradictory values obtained for the Debye temperature fit based on calorimetric and x-ray experiments. Neither boron nor lanthanum atoms behave according to the Debye model: the full dynamics must be taken into account to calculate the MSQD at high and low temperatures.

As a final step of our analysis, we included the phonon effects in the one-step model calculations based on previously developed methodology.[37,38,39,74] The one-step photoemission calculations are depicted in Figure 8b-d, showing the effects of different lattice vibrations theoretical description. The DW factor is implemented with atoms considered "frozen", meaning that lattice vibrations, including zero point motion, are completely neglected (Figure 8b). When T increases, lattice vibrations are taken into account such that phonon-assisted electronic transitions are allowed. In other words, in the photoemission process, the electrons can have access to indirect transitions. Figure 8d shows the case of T = 300 K with an averaged DW factor for both chemical elements. Figure 8c reveals the changes in the spectrum when including species specific DW factors as derived from DFPT calculations. In addition to the ARPES data reported in Figure 6a-b ,Figure SM-1a shows the ARPES image collected at $h\nu$ = 2830 eV and at $T \approx 90$ K. The ARPES images reported in Figures 6a-b and in Figure SM-1a, having similar photon energy but measured at very different temperature, are dramatically different. In Figure SM-1a the bands between 5 eV and 10 eV, mainly attributed to B atoms, are completely smeared out and the only visible dispersing bands are those attributed to La close to the Fermi energy. Furthermore, the ARPES image collected at about 5953.4 eV and $T = 300$ K (Figure SM-1b) shows that B bands are vanished and La bands are almost completely smeared out, leaving a minority of direct transitions, which can barely be extracted from the noise. As shows in Figure 8, at 300 K, the full DFPT phonon DW factor smears the bands significantly more than expected from a simpler model.

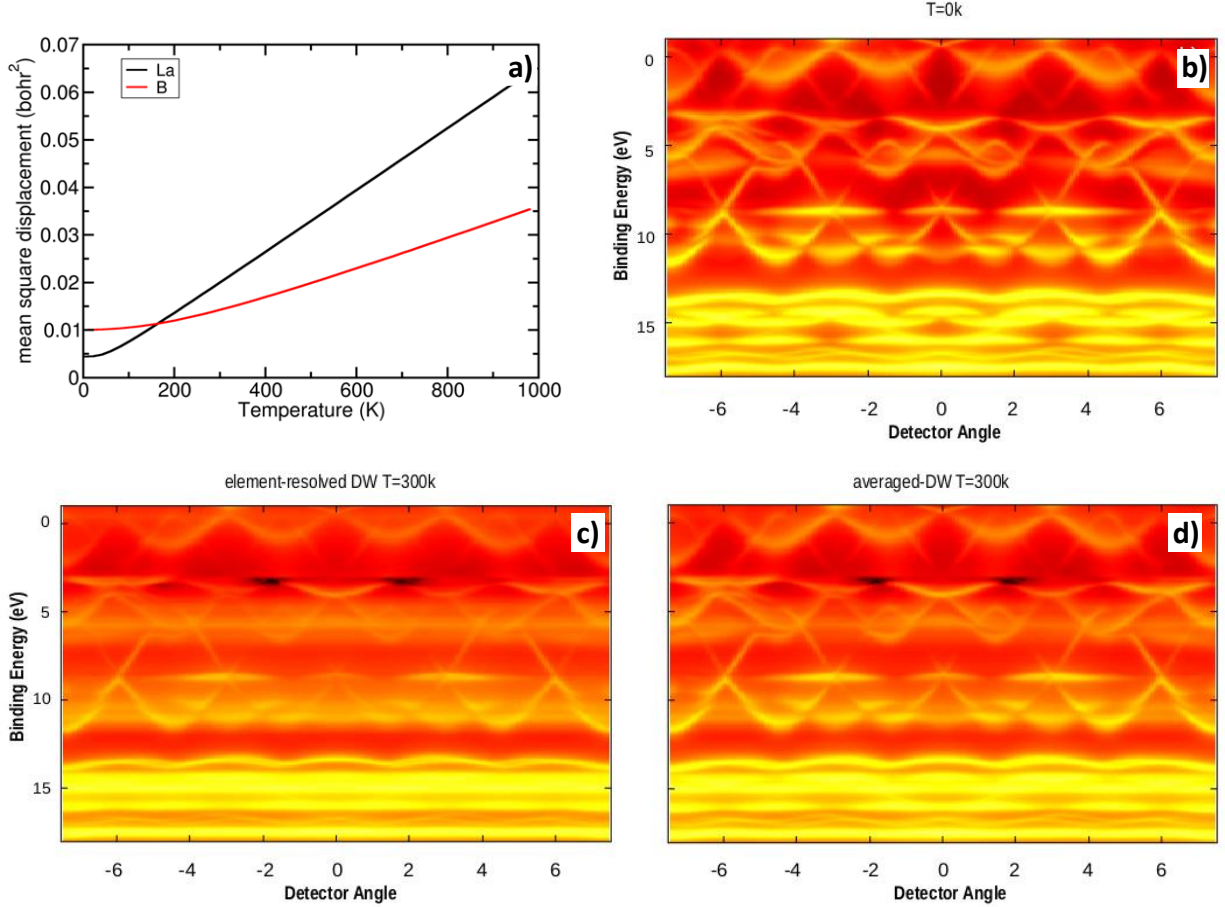

**Figure 8.** a) Mean square displacement of the La and B atoms, as a function of temperature; b) One-step photoemission calculations at $h\nu = 3238$ eV $T = 0$ K; c-d) One-step calculations with different implementation of phonon excitations, either element resolved (c) or averaged (d). The wave-vector blurring is noticeably different, especially for bands around 10 eV, and stronger for the element resolved case.

## V. Conclusion

A detailed characterization of the position and intensity of the XPS core level peaks of the C 1s, O 1s and B 1s and La 3d core levels suggests that the $LaB_6$ crystal studied here does not show bulk impurities and that the carbon and oxygen contaminations are only on the top sample surface. The well-screened and poorly-screened La 3d peaks were identified and interpreted by means of cluster model calculations. We determined that there is a small but non-negligible hybridization between La 4f orbitals and B s-p bands producing the main poorly-screened peaks ($3d^9 4f^0$) at 838 eV and 854.5 eV, respectively, and the clear well-screened satellites ($3d^9 4f^1 \underline{v}$) at 835 eV and at 855.2 eV, respectively. In addition, using tender/hard XPS we demonstrated that, within one crystal, La and B act noticeably differently to the hard x-ray excitation. Their recoil energies are substantially different, specifically it is much larger for B than for La. Interestingly, little to no momentum is transferred to the crystal lattice and B atoms recoil as if they were free on the photoemission timescale.

We also report the first attempted hard X-ray ARPES measurements for a rare-earth hexaboride compound. We have shown that the observed ARPES data can in first approximation be understood within a simple free-electron final-state model. Furthermore, one-step theory provided, at a more quantitative level, a good description of the experimental ARPES data including the relative intensities of most features. The agreement between the experimental data and the one-step model calculations is remarkable.

In the temperature dependence of the experimental data, there is a clear trend for the bands with boron character to broaden and vanish at increasing temperature, due to the breaking of the $\boldsymbol{k}$-conserving dipole selection rule. We model these effects with ab-initio calculations implementing lattice vibrations within the one-step model of photoemission through the Coherent Potential Approximation (CPA), with a qualitative agreement between the trends in experiment and theory. The complex phonon band structure of $LaB_6$ explains the contradictory values obtained for the Debye temperature fit based on calorimetric and x-ray experiments. Neither boron nor lanthanum atoms behave according to the Debye model, dictating a necessity to take into account the full dynamics in order to calculate the atomic mean square displacements (MSQD) at high and low temperatures.

Lastly, we demonstrated that hard X-ray excited ARPES combined with one step-theory of photoemission is a very powerful way experimentally and theoretically to determine true "bulk-like" electronic band structure in complex materials. This successful combination of experimental and theoretical study unlocks a new way of characterizing another class of materials, such as rare-earth hexaborides.

**Supplementary material**

See the supplementary material for: **SM-1** additional ARPES data; **SM-2** relative angular shift of the band structure features with respect to the x-ray photoelectron diffraction pattern; **SM-3** Free Electron Final State calculations; **SM-4** Phonon band structure of $LaB_6$.

**Authors' contributions**

This work is part of A.R. PhD thesis. G.C., C.S.F., A.X.G and S.N. supervised the project. A.R., G.C., S.U., A.X.G and S.N. measured the experimental data. A.R. analyzed all the data. L.N., H.P.M., M.J.V. and J.M. performed theoretical calculations. A.R., G.C., M.J.V., J.M., A.X.G., and S.N. wrote the manuscript. All authors participated in the discussions.

**Data availability**

The data that support the findings of this study are available from the corresponding author upon reasonable request.

**Acknowledgements**

The HXPS measurements at SPring-8 were performed under the approval of NIMS Synchrotron X-ray Station (Proposals No. 2009A4906, No. 2010A4902, and No. 2010B4800), and were partially supported by the Ministry of Education, Culture, Sports, Science and Technology (MEXT), Japan. This research used a HAXPES end-station at beamline 9.3.1 of the Advanced Light Source, LBNL Berkeley (USA), a U.S. DOE Office of Science User Facility under contract no. DE-AC02-05CH11231. A.R. was funded by the Royal Thai Government Scholarship. A.R. acknowledges extra time supported by H. Nakajima. H.P.M. has been supported for salary by the U.S. Department of Energy (DOE) under Contract No. DE-SC0014697. J.M. and L.N. would like to thank the CEDAMNF (CZ.02.1.01/0.0/0.0/15_003/0000358) co-funded by the Ministry of Education, Youth and Sports of Czech Republic. M.J.V. acknowledges funding from ARC project AIMED (Federation Wallonie-Bruxelles G.A. 15/19-09). Computing time was provided by CECI, funded by FRS-FNRS G.A. 2.5020.11 and the Zenobe Tier-1 funded by the Gouvernement Wallon G.A. 1117545 and DECI projects pylight on Beskow and RemEPI on Archer (G.A. 653838 of H2020 and PRACE aisbl). A.X.G. acknowledges support from the U.S. Department of Energy, Office of Science, Office of Basic Energy Sciences, Materials Sciences and Engineering Division under award number DE-SC0019297 during the writing of this paper.

# SUPPLEMENTARY MATERIAL

## SM-1 Additional ARPES data

Figure SM-1a shows clearly La dispersive bands between 0 eV and 4 eV in binding energies, on the other hand, Figure SM-1b shows that, at 6 keV and room temperature, those bands are almost completely smeared out.

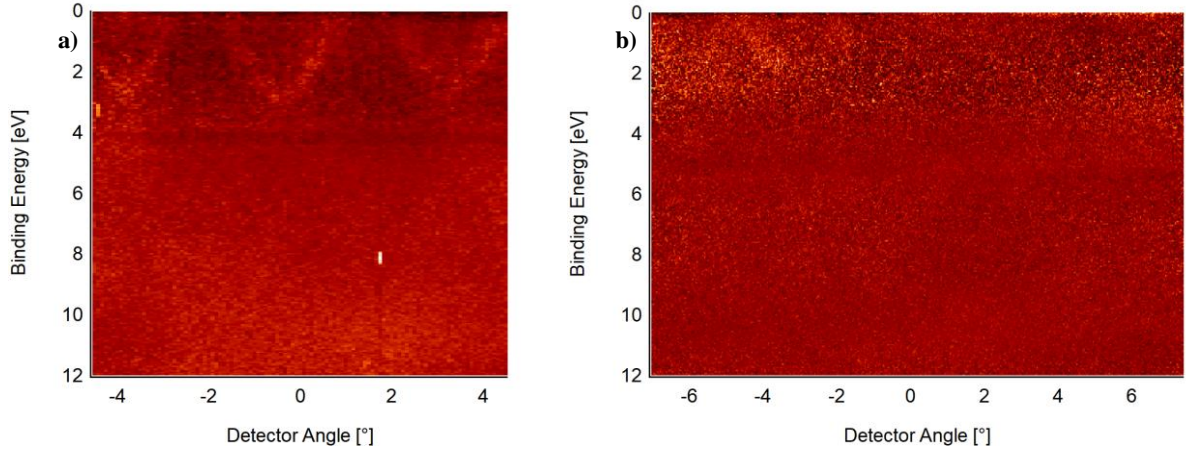

Figure SM-1: ARPES data, normalized by DOS and by XPD modulations: **a)** hv = 2830 eV and T≈ 90 K; **b)** hv = 5953.4 eV and T ≈ 300 K.

## SM-2 Calculated relative angular shift of the band structure features with respect to the x-ray photoelectron diffraction (XPD)

A relative angular shift of the band structure features with respect to the x-ray photoelectron diffraction (XPD) pattern in the detector image occurs due to the fact that in hard x-ray regime the dipole approximation is no longer valid, and the effects of the photon momentum $k_{hv}$ on the magnitude and direction of the final photoelectron momentum $k_f$ must to be considered. This angular deflection is expected to be observed in the position of all band structure features in the detector image. The calculations below are for $hv$=3237.5 eV end $E_{kin}=hv$, which is valid for valence band electrons close to Fermi edge. The magnitude of the final valence-band photoelectron momentum can be computed via

$$\left|\vec{k}_f\right| = 0.512 * E_{kin}^{1/2} = 29.19 A^{-1} \qquad \text{(SM-2a)}$$

The magnitude of the photon momentum is given by

$$\left|\vec{k}_{hv}\right| = 0.0005107 * E_{hv} = 1.64 A^{-1} \qquad \text{(SM-2b)}$$

Since in our experimental geometry the direction of the photoelectron momentum is perpendicular to the direction of the photon momentum, the angular deflection due to the photon momentum vector can be computed using

$$\Delta = \tan^{-1}(1.642/29.19) = 3.219° \quad \text{(SM-2c)}$$

**SM-3 Free Electron Final State calculations (FEFS)**

The FEFS model maps the momentum vector of the photoelectron in vacuum $k_f$ to its initial momentum state in crystal $k_i$ before the photoexcitation. This is based on two assumptions. First, the FEFS model assumes that the momentum components parallel to the sample surface are conserved throughout the entire photoemission process. Second, the momentum component perpendicular to the surface is determined using energy conservation and using the empirical inner potential $V_0$ within the crystal:

$$E_f(k_f) = E_i(k_i) + h\nu = \frac{\hbar^2 k_f^2}{2m_e} - V_0 + \phi_s = \frac{\hbar^2 K^2}{2m_e} + \phi_s \quad \text{(SM-3a)}$$

and

$$k_f = k_i + k_{h\nu} + g_{hkl} \quad \text{(SM-3b)}$$

where $E_f(k_f)$ is the final electronic kinetic energy, $E_i(k_i)$ is the initial energy relative to the Fermi energy, $h\nu$ is the photon energy, $m_e$ is the electron mass, $\phi_s$ is the work function, $k_i$ is the initial-state wave vector within the reduced Brillouin zone (BZ), $k_f$ is the final-state wave vector inside the crystal, $K$ is the wave vector of the photo emitted electron in vacuum, $k_{h\nu}$ is the wave vector of the photon, and $g_{hkl}$ is the bulk reciprocal lattice vector.

Figure 6c) shows a comparison of the FEFS calculation to HARPES data, and even despite some smearing, one can clearly see a good agreement between the calculated "bulk" band structure and the experimental HARPES image.

A particular challenge in HARPES experiments compared to conventional low-energy ARPES is the extreme sensitivity to the sample alignment. Due to the extreme length of the photoelectron momentum vector $k_f$ at photon energies of several keV, even misalignments in the order of 0.1° result in a sizable displacement within the reduced Brillouin zone scheme. Even though the sample was carefully aligned, we had to vary the sample geometry that was used for our calculations, and we found the best agreement after tilting the sample by 0.5° compared to the initially assumed sample geometry.[73]

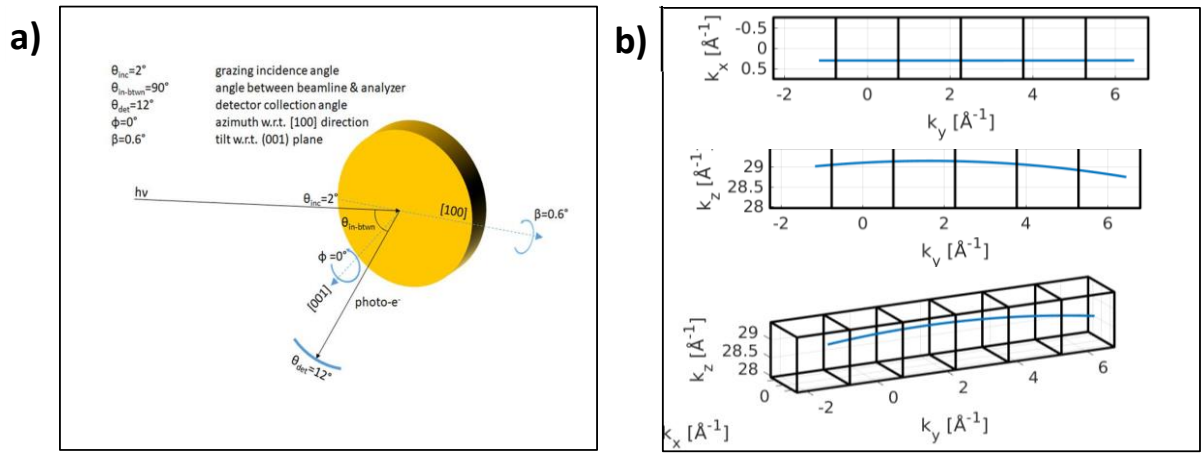

Figure SM-3: a) Geometry used in FEFS calculations; b) HARPES paths in k-space for the optimized experimental geometries determined from the FEFS calculations. The paths are shown in three different perspectives.

## SM-4 Figures of a) Phonon band structure of LaB$_6$ and b) Specific heat of LaB$_6$

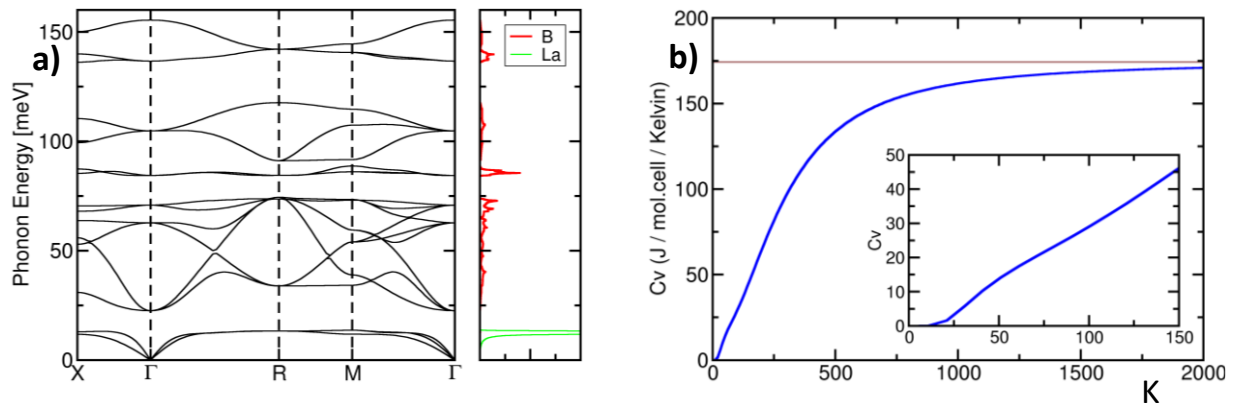

Figure SM-4a) Phonon band structure of LaB$_6$, with element projected Density of States in side panel. SM-4b) Specific heat of LaB$_6$ within the harmonic approximation and DFPT, as a function of temperature in Kelvin [K]. Inset shows the small bump created by the Lanthanum manifold around 50 K.